\documentstyle[multicol,aps,epsf]{revtex}

\begin{document}
\title{Molecular Simulations of Dewetting}
\author{Joel Koplik}
\address{Benjamin Levich Institute and Department of Physics,
City College of the City University of New York, New York, NY 10031}
\author{Jayanth R. Banavar}
\address{Department of Physics, 
The Pennsylvania State University, University Park, PA 16802}
\date{\today}
\maketitle

\begin{abstract}
We have studied the breakup and subsequent fluid flow in very thin films of 
partially wetting liquid on solid substrates, using molecular dynamics 
simulations.  The liquid is made of short chain molecules interacting with
Lennard-Jones interactions, and the solid is modeled as a clean crystal 
lattice whose atoms have thermal oscillations.  Films below a critical 
thickness are found to exhibit a spontaneous spinodal-like instability
leading to dry patches, as predicted theoretically and observed in some
experiments.  Liquid withdrawing from a dry patch collects in a moving rim 
whose fluid dynamics is only partially in agreement with earlier predictions.

\end{abstract}

\begin{multicols}{2}
The equilibrium configuration of a partially wetting liquid on a solid
substrate is a drop, with finite contact angle given by Young's equation.
If the liquid is initially placed on the substrate in the form of a uniform
coating film, however, it is potentially unstable and may ``dewet'' the 
solid so as to contract into drops, or perhaps other shapes.  This process 
\cite{bd,reiter} is obviously 
very relevant to applications in materials containment and transport, and 
at the same time involves a number of challenging
scientific issues, such as the nature of the instability, the fluid dynamics
of the dewetting liquid, and the shape selection problem for the final pattern.
In this Letter, we describe the results of molecular dynamics (MD)
simulations which address the first two issues.  We first present new and
independent confirmation of the spinodal dewetting scenario for very thin 
films \cite{spin,bmq}, those of sub-micron thickness where gravity is
negligible but Van der Waals interactions between the liquid and the solid 
are crucial.  We observe unstable
films whose behavior reflects spontaneous rather than 
nucleated instabilities.  We then consider the hydrodynamics of a liquid
film undergoing a dewetting flow, and determine the time evolution of the
film length and shape, and the issues of internal flow and contact angle 
variability.  Here we find only partial agreement with earlier theoretical
proposals.  The pattern selection issue unfortunately involves the dynamics 
on larger scales than those readily addressed with MD, and is not considered
here.  Several previous papers have used molecular simulations to study 
dewetting \cite{mol}, but with an emphasis on different parameters and
aspects of the
instability, the behavior at shorter times than we consider, and no
consideration of the hydrodynamics of the dewetting liquid.  Furthermore,
we disagree with some of their conclusions.
 
The behavior of a thin liquid film on a solid depends on Van der Waals (VdW)
forces \cite{isr} if the film is sufficiently thin, gravity otherwise, surface 
tension if its surface is not flat, and when motion occurs, viscosity.  In 
the partially wetting case, the VdW forces favor a solid exposed to vapor 
rather than liquid, but for incompressible liquids any attempt to withdraw
the liquid increases the liquid-vapor surface area.  The competition between 
VdW and surface tension forces
forces leads to an instability criterion that small sinusoidal perturbations 
of wavelength greater than $\lambda_c= \sqrt{4/3} \pi h_0^2 /a$ 
are unstable \cite{bd} . Here $a$ is a microscopic length given by $a^2 =
(A_{LL}-A_{LS})/6\pi\gamma$, where the $A_{ij}$ ($i,j$ = liquid L, solid 
S or vapor V) are the Hamaker constants, $\gamma$ is the LV surface tension, 
$\mu$ is the liquid viscosity, and $h_0$ is the unperturbed thickness. 
The linear stability calculation shows that there is a most unstable 
wavelength $\sqrt{2}\lambda_c$ with corresponding growth rate 
$w_{\rm max} = 3a^4\gamma / \mu h_0^5$.  Further analysis, using
weakly nonlinear theory \cite{nl} or lubrication modeling \cite{lub} 
may then be used to discuss the final patterns.  A long-standing issue has 
been whether the
origin of observed dewetting phenomena is simply the evolution of these
``spontaneous'' instabilities, analogous to spinodal decomposition, or 
whether nucleation at impurity sites or
by contaminants is required.  Recent experiments and previous simulations
favor the former mechanism, and we wish to provide further evidence, by
systematically varying the film thickness.  The procedure will be to
simulate a clean substrate of width $L$ with periodic boundary
conditions, and films with various values of $h_0$.  The maximum wavelength 
of any disturbance is on the order of magnitude of the substrate size, so an 
instability should occur only when $\lambda_c(h_0)$ is less than O($L$).

Our simulations are based on standard MD techniques applied to fluid flows 
\cite{at,arfm}, and the details are very similar to that used in our earlier
studies of drop wetting \cite{spread}.  We consider a system with 18000
fluid molecules consisting of four atom FENE \cite{fene} chains, placed on a
substrate in the form of an fcc-crystal with atoms tethered by linear
springs to regular lattice sites.  
All atoms interact via two-body Lennard-Jones interactions, 
$V(r)=4\epsilon [\, (r/\sigma)^{-12}-c_{ij}(r/\sigma)^{-6}\, ]$, 
where $i,j$ refer to the fluid and solid species present.  The potential is
cutoff at 2.5$\sigma$ for computational speed, but a $z^{-3}$ tail is added
to the solid-liquid interaction above the cutoff distance, 
corresponding to the interaction due to a half-space of solid.  
The temperature is 1.0, maintained by
a Nos\'e-Hoover thermostat, and the fluid density is 0.8$\sigma^{-3}$.
The solid has 34848 atoms, in the shape of a square substrate of side 
102.6$\sigma$, and two fcc cells (four staggered layers) in thickness.  
We prefer to use molecular chains rather than a monatomic liquid so as to
reduce the vapor
pressure and have a sharper interface.  FENE molecules have non-Newtonian 
behavior under high-stress conditions, but here the stress is usually weak 
(but see below) and any other potential confining atoms into molecules
would suffice.  Note that the $r^{-6}$ term in the
potential and the $z^{-3}$ tail are the microscopic basis of the 
semi-macroscopic VdW force, and the wettability of the solid is controlled 
by the Lennard-Jones solid-liquid 
interaction coefficient $c_{LS}$.  By simulating the spreading of a liquid
drop on this substrate, we find that $c_{LS}=1$ gives complete wetting,
while $c_{LS}=0.75$ gives partial wetting. 
We observe in these microscopic simulations that the
wetting and dewetting behavior is essentially the same if we remove the 
$z^{-3}$ tail but increase $c_{LS}$.  Similarly, although we prefer the use
of a solid made of tethered molecules which can realistically exchange
energy with the liquid, the dewetting behavior is not significantly modified
if fixed potential sites are used.  The in-plane variation of the wall
interaction is crucial, however, because if the potential is purely a function
of height, the liquid readily slips along the substrate.

\begin{figure}
\narrowtext
\epsfxsize=3.0in\epsfysize=3.0in
\hskip 0.01in\epsfbox{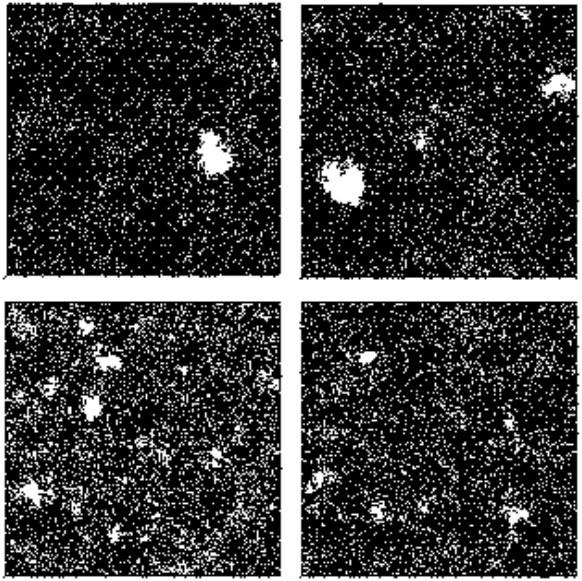}
\vskip 0.1in
\caption{Initial stages of dewetting for various thicknesses.
Clockwise from upper left, the initial thicknesses and elapsed times
are, respectively, (6.45$\sigma$,300$\tau$), (5.7$\sigma$,250$\tau$),
(5.20$\sigma$,100$\tau$), (4.95$\sigma$,25$\tau$).}
\end{figure}

We have simulated the free evolution of uniform liquid films of various 
initial thicknesses, and Fig.~1 shows snapshots for four choices near the 
onset of instability.  Each frame is a top view of a liquid film, where a
molecule is displayed as the three-segment broken line joining its four 
atomic centers, so that darkness increases with thickness.  The initial film 
is slightly
thicker than those shown, $h_0$= 6.7$\sigma$, and is stable over a time 
1200$\tau$, where $\tau=\sigma\sqrt{\epsilon/m}$ is a characteristic molecular
time (in the picosecond range).  Thinner films, produced by skimming
molecules off the surface of the liquid, are unstable;  the different
frames of the figure show films at an early stage of evolution just after
one or more holes appear, corresponding to dry patches on the substrate.
The solid is flat, aside from thermal fluctuations in its atomic positions, 
and the liquid is homogeneous, so there is no nucleation.  Furthermore, the
locations of the dry spots for various films are uncorrelated.  Once a dry spot
appears, it invariably continues to grow, and eventually the liquid withdraws
into a single round drop.  Liu et al. \cite{mol} claim to the contrary to
observe a nucleation mechanism, wherein dry patches smaller than some critical 
size heal.  We find instead that once a dry region larger than a couple of
molecular diameters appears, which is to say once the molecules on the two
sides are out of mutual interaction range, it continues to grow.

The effect of varying the thickness in the
unstable cases is that decreasing $h_0$ decreases the time before a dry spot
appears, and also decreases the spacing between the spots.  The stability
calculation summarized above predicts that these two quantities would vary as
$h_0^5$ and $h_0^2$, respectively.  It would be prohibitive in computation
time to test these results quantitatively, but the expected trend is clearly
reproduced.  A further difficulty in making a quantitative comparison 
is the unknown connection between the atomic interaction parameters and the 
Hamaker constants.  There is a standard relation \cite{isr} for a monatomic 
liquid neglecting thermal motion, $A_{ij}=\epsilon\sigma^6\pi^2\rho_i\rho_j 
c_{ij}$ where $\rho_i$ is the density of species $i$,
but not for molecules.  If we use this expression anyway, along with
measurements of $\gamma$ and $\mu$ from earlier MD simulations, our results 
agree with the stability calculations to within factors of 2-3.   

Now we turn to consideration of the fluid mechanics of the withdrawing liquid.
Conventional theoretical and numerical studies of dewetting flows are plagued 
by several difficulties.  In general, this is a moving boundary 
problem where disjoining pressure must be included to take account of VdW 
forces.  Full Stokes equation calculations are difficult due to the 
fine resolution that would be required near the solid to capture this term.
Furthermore, there is a moving contact line, with the usual accompanying 
shear stress singularity \cite{sing}, and where one is completely ignorant 
of the appropriate {\em dynamic} contact angle.  The singularity may be 
eliminated by use of a slip model for the boundary value of the fluids 
velocity \cite{slip}, but the correct angle and its possible variation 
with flow conditions is simply unknown.  The computational complexity may
be significantly reduced by use of a depth-averaged or lubrication model 
\cite{lub}, but the standard parameterization of disjoining pressure, as 
a single term proportional to the inverse square of the thickness, does not 
lead to an actual dewetting motion, but rather pins the edge of the film.  
This deficiency can be remedied by modifications to the potential which give
a minimum at a small value of thickness, but in this case a residual film is 
left behind, not usually seen in experiment.

The only quantitative theoretical attempt to consider the fluid mechanics of 
dewetting, by Brochard and collaborators \cite{bmq}, is based on the strong 
simplifying assumptions that the liquid collects into a growing rim as it 
dewets with negligible fluid motion elsewhere, that the rim is in the shape 
of an arc of a circle, and that particular contact angles 
appear at the two edges of the rim.  It is then possible to
derive power laws for the speed and shape evolution of the rim.  
Of the assumptions made in this treatment, it is entirely reasonable that the 
liquid in the center of the film would be static until it is swept
up by a withdrawing rim, but the particular assumptions made about the shape 
of the rim are less compelling.  The theoretically unbiased MD simulations 
reported here provide new information at microscopic resolution.  

\begin{figure}
\narrowtext
\epsfxsize=3.0in\epsfysize=3.0in
\hskip 0.01in\epsfbox{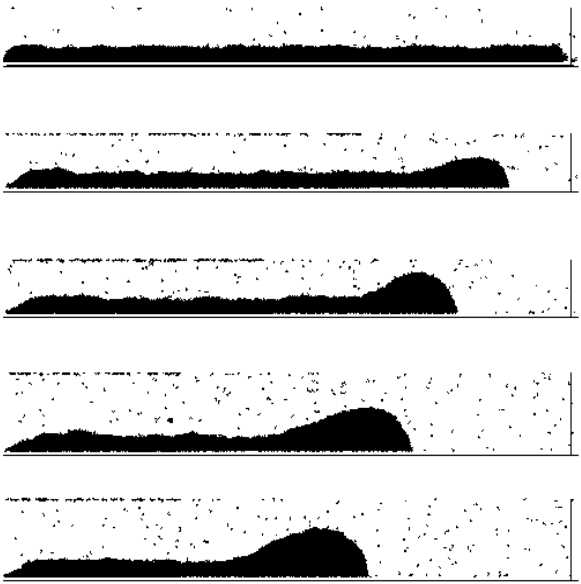}
\vskip 0.1in
\caption{Evolution of a dewetting film, seen from the side, at times 
100$\tau$, 1250$\tau$, 2500$\tau$, 3750$\tau$ and 5000$\tau$.}
\end{figure}

As in the previous calculations, we begin with a solid substrate 
initially covered by a thin uniform liquid film, with the same density and
temperature as above, but thickness 12$\sigma$.  In order to maximize the
time and space interval over which the film evolves, we take the 
substrate to be a long narrow strip, 547$\sigma$ x 17$\sigma$, so as to 
discourage instabilities in the spanwise direction.  The liquid film is first 
equilibrated with periodic boundary conditions and with $c_{LS}$=1, 
corresponding to complete wetting.  A strip of liquid is removed ``by hand'' 
at one edge of the film while $c_{LS}$ is reduced to a dewetting value 0.75 
for the solid atoms in the right-hand quarter of the 
substrate where motion occurs.  The film at this stage is shown from 
the side in Fig.~2a, and in subsequent frames the liquid at the edge of 
the film withdraws from the substate and collects in
a rim which translates and grows, at least initially.  As the film dewets,
we occasionally extend the region of substrate where $c_{LS}=0.75$ to keep
it ahead  of the advancing rim, while retaining the ``wetting value'' 1.0 
elsewhere.  In this way, the motion 
is purely from right to left, rather than inwards on both sides, and it is 
possible to study the behavior of a longer dewetted region. 

\begin{figure}
\narrowtext
\epsfxsize=3.0in\epsfysize=3.0in
\hskip 0.01in\epsfbox{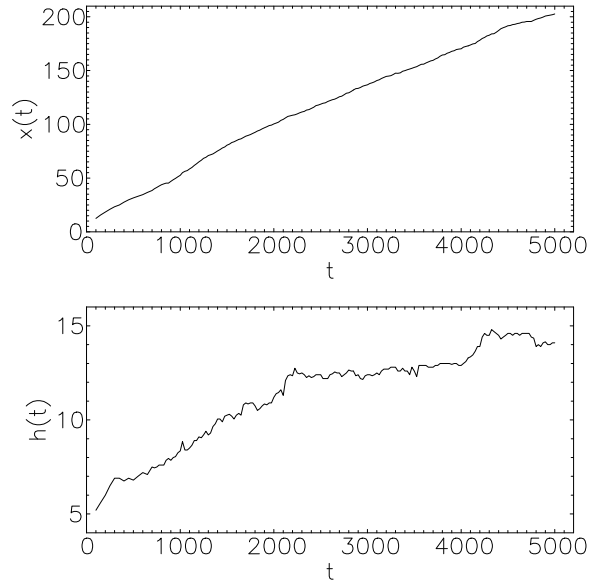}
\vskip 0.1in
\caption{Time-dependence of the length (top) and rim height (bottom) of the 
dewetting film shown in Fig.~2.  The units are $\sigma$ in the ordinates
and $\tau$ in the abscissa.}
\end{figure}

We see that while the dewetted liquid indeed forms a rim which grows
in size as it moves, only the outer edge is circular, while the inner 
boundary is more ramp-like.  There is no second contact angle \cite{bmq}
where the rim meets the flat part of the film.  The solid-liquid-vapor 
contact angle at the receding edge is seen 
to fluctuate with time, as usual at this scale, but its mean value is
observed to be roughly 90$^\circ$ and is distinctly less than the static 
equilibrium value of 105$^o \pm 5^\circ$.  Hwang et al. \cite{mol} present 
plots of contact angle {\em vs}. time which suggest a smooth and systematic 
variation, but in our simulations (with larger scale and different
parameters) substantial oscillation is seen.  The position of the edge of the 
film as a function of time is 
shown in Fig.~3, and as predicted \cite{bmq} its speed is roughly constant.
The glitch around time 1000$\tau$ is related to the first appearance of an 
obvious rim, and the possible flattening at the largest times is probably
due to gradual thickening of the immobile part of the film.  Liu et al.
\cite{mol} however find a different behavior at early times in other systems.
The height and length of the rim were predicted theoretically to vary as 
$\sqrt{t}$, at least for non-volatile liquids, but we observe a much slower 
growth of the rim.  The rim is subject to avalanching along its inner side,
and grows irregularly at a rate very roughly $t^{1/4}$, Fig.~3, while 
the width of the rim region appears to stabilize.  
Additional, more microscopic, information is provided by the simulation as
well.  Typical velocity fields are shown in the rim region
and in the middle of the film, respectively, in Fig.~4.  There is slip
in the very near vicinity of the contact line but not elsewhere.  The shear
stress and pressure do not show any statistically significant variation,
aside from a shear stress peak at the contact line which usually accompanies
slip there \cite{slip}.  The flow is essentially confined to the rim region, 
as predicted, 
but the details of the velocity field are hard to resolve because the values 
are small and subject to small-system thermal fluctuations.  While there
is little bulk motion in the plateau region of the film, there is some 
motion in the vapor and also along the surface however;  the snapshots in 
Fig.~2 show some wave-like excitations on the film surface, and there is 
a non-zero velocity at the top of the liquid plateau indicated in Fig.~4.  

\begin{figure}
\narrowtext
\epsfxsize=3.0in\epsfysize=3.0in
\hskip 0.01in\epsfbox{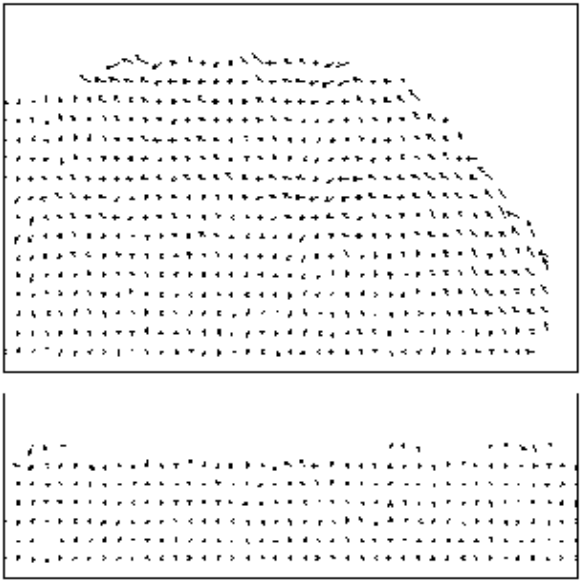}
\vskip 0.1in
\caption{Velocity fields at the edge (top) and in the middle (bottom) of 
a receding film, averaged over the time interval 3125-3150$\tau$ and spatial 
bins of cross section 1.71$\sigma$.  The largest velocity vector shown is 
0.2$\sigma /\tau$.}
\end{figure}

Although in this Letter we have focused on the results of just two 
rather large 
scale simulation sequences, we emphasize that numerous smaller-sized 
runs with various alternative parameters exhibit essentially the same 
behavior.  Among the variations we have explored in this way are different
molecular sizes and bonding potentials, different substrate thicknesses, 
other values of the temperature, adding a (weak) gravitational force, 
and different methods for initialization.  In particular, aside from the 
obvious expectation that decreasing the strength of the solid-fluid 
interaction enhances the rate of dewetting, we find that 
sufficiently thin dewetting films always follow the trends expected from
the spinodal decomposition scenario and its linear stability analysis, and
that dewetted fluid collects from the substrate into a rim moving with roughly 
constant velocity.  The detailed fluid dynamics of flow in the rim region 
and the shape evolution of the front is unfortunately not yet predictable. 
A particularly fruitful area for further study might be
the effects of molecular structure and interactions on dewetting dynamics.
The advantage of having the forces and structures under explicit control
may help in the design of controlled dewetting patterns.  

\medskip
We thank M. Fermigier and L. Limat for instigating our interest in this
problem, and M. Miksis, L. W. Schwartz and S. Troian for helpful discussions.
This research was supported by the NASA Microgravity Science and 
Applications Division, and computer time was provided by the NASA Center for
Computational Sciences and the San Diego Supercomputer Center.

\end{multicols}

\end{document}